\RequirePackage[mathlines]{lineno} 
\documentclass[aps,prd,twocolumn,showpacs,amsmath,amssymb]{revtex4-1}
\usepackage{epsfig}
\usepackage{graphicx}
\usepackage{dcolumn}
\usepackage{bm}
\usepackage{overpic}
\usepackage{subfigure}
\usepackage{float}
\usepackage{color}
\usepackage{amsmath}
\usepackage{mathcomp}
\usepackage{multirow}
\usepackage[dvipdfm, colorlinks,,linkcolor=blue, citecolor=blue, urlcolor=green]{hyperref}
\usepackage{rotating}

\newcommand{\chicJ}{\chi_{cJ}}

\newcommand{\ppp}{\pi^+\pi^- \pi^0}

\newcommand{\GG}{\gamma\gamma}

\newcommand{\kk}{K^+K^-}

\newcommand{\psp}{\psi(3686)}

\begin{document}
\normalsize
\parskip=5pt plus 1pt minus 1pt

\title{ \boldmath Observation of OZI suppressed decays $\chi_{cJ}\to\omega\phi$}

\author{
  \begin{small}
    \begin{center}
      M.~Ablikim$^{1}$, M.~N.~Achasov$^{10,d}$, S. ~Ahmed$^{15}$, M.~Albrecht$^{4}$, M.~Alekseev$^{55A,55C}$, A.~Amoroso$^{55A,55C}$, F.~F.~An$^{1}$, Q.~An$^{42,52}$, J.~Z.~Bai$^{1}$, Y.~Bai$^{41}$
% ...
      \\
      \vspace{0.2cm}
      (BESIII Collaboration)\\
      \vspace{0.2cm} {\it
        $^{1}$ Institute of High Energy Physics, Beijing 100049, People's Republic of China\\
% ...
$^{4}$ Bochum Ruhr-University, D-44780 Bochum, Germany\\
% ...
$^{10}$ G.I. Budker Institute of Nuclear Physics SB RAS (BINP), Novosibirsk 630090, Russia\\
% ...
$^{15}$ Helmholtz Institute Mainz, Johann-Joachim-Becher-Weg 45, D-55099 Mainz, Germany\\
% ...
$^{41}$ Southeast University, Nanjing 211100, People's Republic of China\\
$^{42}$ State Key Laboratory of Particle Detection and Electronics, Beijing 100049, Hefei 230026, People's Republic of China\\
% ...
$^{52}$ University of Science and Technology of China, Hefei 230026, People's Republic of China\\
% ...
$^{55}$ (A)University of Turin, I-10125, Turin, Italy; (B)University of Eastern Piedmont, I-15121, Alessandria, Italy; (C)INFN, I-10125, Turin, Italy\\
% ...
$^{d}$ Also at the Novosibirsk State University, Novosibirsk, 630090, Russia\\
% ...
      }
    \end{center}
    \vspace{2cm}
  \end{small}
}

\affiliation{}


\vspace{4cm}

\date{\today}

\begin{abstract}
  Decays $\chi_{cJ}~(J=0,1,2)\to\omega\phi$ are studied using $(448.1\pm2.9)\times 10^{6} ~\psi(3686)$ events collected
  with the BESIII detector in 2009 and 2012.
  In addition to the previously established
$\chi_{c0}\to\omega\phi$, first observation of
$\chi_{c1} \to \omega \phi$ is reported in this paper. The measured
product branching fractions are
  ${\cal{B}}(\psi(3686)\to\gamma\chi_{c0})\times{\cal{B}}(\chi_{c0}\to\omega\phi)=(13.83\pm 0.70\pm 1.01)\times10^{-6}$ and
  ${\cal{B}}(\psi(3686)\to\gamma\chi_{c1})\times{\cal{B}}(\chi_{c1}\to\omega\phi)=(2.67\pm 0.31\pm
  0.27)\times10^{-6}$, and the absolute branching fractions are ${\cal{B}}(\chi_{c0}\to\omega\phi)=(13.84\pm 0.70\pm 1.08)\times10^{-5}$ and
  ${\cal{B}}(\chi_{c1}\to\omega\phi)=(2.80\pm 0.32\pm
  0.30)\times10^{-5}$.
  We also find a strong evidence for $\chi_{
    c2}\to\omega\phi$ with a statistical significance of 4.8$\sigma$, and the
  corresponding product and absolute branching fractions are measured to be
  ${\cal{B}}(\psi(3686)\to\gamma\chi_{c2})\times{\cal{B}}(\chi_{c2}\to\omega\phi)=(0.91\pm0.23\pm0.12)\times10^{-6} $ and ${\cal{B}}(\chi_{c2}\to\omega\phi)=(1.00\pm0.25\pm0.14)\times10^{-5}$. Here, the first errors are statistical and the second ones systematic.
\end{abstract}


\maketitle

\section{Introduction}
\label{sec:introduction}
\vspace{-0.4cm}

 The lowest triplet $P$-wave states of charmonium (the $c\bar{c}$ bound state),
 ${\chi}_{cJ}(1P)$, with quantum numbers $I^{G}J^{PC}$ $=$ $0^{+}J^{++}$
 and $J$ $=$ $0$, $1$, and $2$, can be found abundantly in the electromagnetic
 decays ${\psi}(3686)$ ${\to}$ ${\gamma}{\chi}_{cJ}$ with an approximate branching
 fraction of $30\%$ \cite{pdg}.
 The ${\psi}(3686)$ meson can be directly produced at the $e^{+}e^{-}$ colliders,
 such as the BEPCII \cite{nima614.345}, where the ${\chi}_{cJ}$ meson
 are easily accessible by the electromagnetic
 decays ${\psi}(3686)$ ${\to}$ ${\gamma}{\chi}_{cJ}$.

 The hadronic ${\chi}_{cJ}$ decays are important probes of the strong force
 dynamics. First of all, the mass of the $c$ quark (${\sim}$ $1.5$ GeV$/c^{2}$) is
  well known between the perturbative and nonperturbative
 QCD domains in theoretical calculations.
 Due to the complexity and entanglement of the long- and short-distance
 contributions, 
 large theoretical uncertainties of branching ratios for the ${\chi}_{cJ}$ ${\to}$ $VV$ decays are known \cite{pan70.53,prd72.094018,
 plb611.123,prd72.074001,plb659.221,prd81.014017,prd81.074006}. (In this
 paper, the symbol of $V$ denotes the ${\omega}$ and ${\phi}$ mesons).
 The hadronic ${\chi}_{cJ}$ decays provide a prospective laboratory to
 limit theoretical parameters and test various phenomenological models.
 Second, the ${\chi}_{cJ}$ mesons have the same quantum numbers
 $J^{PC}$ as some glueballs and hybrids, although none of the glueball and
 hybrid states has been seen until now\cite{glueball}.
 The hadronic ${\chi}_{cJ}$ ${\to}$ $VV$ decays are ideal objects to exploit
 the glueball-$q\bar{q}$ mixing and the quark-gluon coupling of the strong
 interactions at the relatively low energies.
 Third, the ${\chi}_{cJ}$ mesons are below the open-charm threshold.
 Most of the hadronic ${\chi}_{cJ}$ decay modes are suppressed by the
 Okubo-Zweig-Iizuka (OZI) rule \cite{ozi}. It is shown in the previous
 theoretical researches that the contributions from the intermediate
 glueballs or hadronic loops can scuttle the OZI rule in the ${\chi}_{cJ}$
 ${\to}$ $VV$ decays \cite{prd47.5050,prd64.094507,prd53.6693,plb631.22},
 and avoid the so-called helicity selection (HS) rule (also called the ``naturalness''
 which is defined as ${\sigma}$ $=$ $(-1)^{S}P$ \cite{npb201.492},
 where $S$ and $P$ are respectively the spin and parity of the particle.)
 in the ${\chi}_{c1}$ ${\to}$ $VV$ decays \cite{prd81.014017,prd81.074006}.

 The ${\chi}_{cJ}$, ${\phi}$ and ${\omega}$ mesons
 differ from each other in their quark components according to the quark model
 assignments. This fact causes the ${\chi}_{cJ}$ ${\to}$ ${\omega}{\phi}$
 decay modes to be doubly OZI (DOZI) suppressed, and results in the branching
 fractions for the ${\chi}_{cJ}$ ${\to}$ ${\omega}{\phi}$ decays much less than
 those for the singly OZI-suppressed ${\chi}_{cJ}$ ${\to}$ ${\omega}{\omega}$,
 ${\phi}{\phi}$ decays \cite{pdg,prl107.092001}.
 In reality $\omega$ and $\phi$ are not ideal mixtures of the flavor SU(3) octet and singlet \cite{prd59. 114027},
 which would provide a source that violates the DOZI suppressed rule for ${\chi}_{c1}$ ${\to}$ ${\omega}{\phi}$.
 The DOZI-suppressed ${\chi}_{cJ}$ ${\to}$ ${\omega}{\phi}$ decays have been
 observed based on the $106{\times}10^{6}$ ${\psi}(3686)$ events accumulated
 with the BESIII detector in 2009, with significances of $10\,{\sigma}$,
 $4.1\,{\sigma}$ and $1.5\,{\sigma}$ for the ${\chi}_{c0}$, ${\chi}_{c1}$ and
 ${\chi}_{c2}$ decays, respectively \cite{prl107.092001}.

 In this paper, the ${\chi}_{cJ}$ ${\to}$ ${\omega}{\phi}$ decays will be
 reinvestigated via the radiative transitions ${\psi}(3686)$ ${\to}$ ${\gamma}{\chi}_{cJ}$
 with combined experimental data, {\em i.e.}, $(448.1{\pm}2.9){\times}10^{6}$
 ${\psi}(3686)$ events collected with the BESIII detector during 2009 and 2012
 \cite{cpc42.023001}.

\section{BESIII detector and monte carlo simulation}
\label{sec:BESIII}
\vspace{-0.4cm} The BESIII detector  operating at the BEPCII collider
is described in detail in
Ref.~\cite{nima614.345}.  The detector is cylindrically symmetric and
covers 93\% of $4\pi$ solid angle.  It consists of the following four
sub-detectors: a 43-layer main drift chamber (MDC), which is used to determine
momentum of the charged tracks with a resolution of 0.5\% at 1 GeV/$c$
in the axial magnetic field of 1 T; a plastic scintillator
time-of-flight system (TOF), with a time resolution of 80~ps (110~ps)
in the barrel (endcaps); an electromagnetic calorimeter (EMC)
consisting of 6240 CsI(Tl) crystals, with photon energy resolution at
1 GeV of 2.5\% (5\%) in the barrel (endcaps); and a muon counter
consisting of 9~(8) layers of resistive plate chambers in the barrel
(endcaps), with position resolution of 2 cm.

The \textsc{geant}4-based~\cite{numa506.250,tns53.270} Monte Carlo (MC)
simulation software \textsc{boost}~\cite{cpc30.371} includes the geometry and
material description of the BESIII detectors, the detector response
and digitization models, as well as a database that keeps track of 
of the
running conditions and the detector performance.
MC samples are used to optimize
the selection criteria, evaluate the signal efficiency, and
estimate physics backgrounds.  An inclusive MC sample of
$5.1\times10^8$ $\psi(3686)$ events is used for the background studies.  The
$\psi(3686)$ resonance is produced by the event generator \textsc{kkmc}~\cite{kkmc}, where the initial state radiation is
included, and the decays are simulated by {\sc evtgen}~\cite{evtgen} with
known branching fractions taken from Ref.~\cite{pdg},
while the unmeasured decays are
generated according to \textsc{lundcharm}~\cite{lundcharm}.  The signal
is simulated with the decay $\psi(3686)\to\gamma\chi_{cJ}$
generated assuming an electric-pole ($E1$) transition.
 The decay $\chi_{cJ}\to\omega\phi$ is generated using \textsc{helamp}~\cite{evtgen}, the helicity amplitude model where the angular correlation between $\omega$ decay and $\phi$ decay has been considered. Ref.~\cite{prl107.092001} shows that
the model describes the experimental angular distribution well.
We assume $\chi_{ cJ}\to\omega\phi$ and $\chi_{cJ}\to\phi\phi$
have the same helicity amplitudes with the same \textsc{helamp} parameters. In addition, $\chi_{cJ}$ states are simulated using a relativistic Breit-Wigner incorporated within the helicity amplitudes in the {\sc evtgen} package~\cite{evtgen}. The background
decays $\chicJ\to\omega\kk$, $\phi\ppp$, and the nonresonant decay $\chicJ\to\kk\ppp$ are generated using the phase space model.

\section{Event Selection}
\label{sec:selection}
 In this analysis, the $\phi$ mesons are reconstructed by
$\kk$, while $\omega$ by $\ppp$. Event candidates are required to have four well-reconstructed tracks from charged particles with zero net charge, and at least three good photon candidates.

A charged track reconstructed from MDC hits
should have the polar angle, $\theta$,
$|\rm cos\theta|<0.93$  and pass within $\pm$10 cm of the interaction point along
the beam direction and within 1 cm in the plane perpendicular to the
beam.  To separate $K^\pm$ from $\pi^\pm$, we require that at least
one track is identified as a kaon using $dE/dx$ and TOF information.
If the identified kaon has a positive (negative) charge,
 the second kaon is found by searching for a combination that
minimizes
 $|M_{K^+K^-}-M_{\phi}|$, among all identified kaons and the
negative (positive) charged tracks, where $M_{K^+K^-}$ is the invariant mass of the identified kaon and an unidentified track
with kaon mass hypothesis,
and $M_{\phi}$ is the nominal $\phi$ mass~\cite{pdg}. The remaining two charged tracks are assumed to be pions.

The photon energy deposit is required to be at least
25 MeV in the barrel region of the EMC ($|\rm cos\theta|< 0.80$) or
50 MeV in the EMC endcaps ($0.86 < |\rm cos\theta| < 0.92$). To suppress electronic noise and energy deposits
unrelated to the event, the EMC time $t$ of the photon candidates must
be in coincidence with collision events within the range ${0}\leq t\leq 700$ ns.
At least three photons are required in an event.

In order to improve the mass resolution, a four-constraint ($\rm 4C$) kinematic  fit is performed by assuming
energy-momentum conservation for the
$\psi(3686)\to 3\gamma K^+ K^-\pi^+\pi^-$ process.
If the number of photons is larger than three, then looping all $3\gamma K^+K^-\pi^+\pi^-$ combinations and the one with the smallest $\chi^{2}_{4C}$ is chosen.
The event is kept for further analysis if
$\chi^{2}_{4C}(3\gamma K^+K^-\pi^+\pi^-)<60$, which is obtained by optimizing the figure of merit (FOM) $S/\sqrt{S+B}$, where $S$ and $B$ are the numbers of MC simulated signal and background events, respectively. In addition, $\chi^{2}_{4C}(3\gamma K^{+}K^{-}\pi^{+}\pi^{-})$ $<$
$\chi^{2}_{4C}(4\gamma K^{+}K^{-}\pi^{+}\pi^{-})$ is applied to suppressed the background with an extra photon in the final state.

A further requirement of $|M^{\rm recoil}_{\pi^+\pi^-}-M_{J/\psi}|>8$ MeV/$c^2$ obtained by optimizing FOM, is applied to suppress the $\psi(3686)\to\pi^+\pi^-J/\psi$ background,
where the $M^{\rm recoil}_{\pi^+\pi^-}$ is the recoil mass
for the $\pi^+\pi^-$ system, and $M_{J/\psi}$ is the nominal mass of $J/\psi$ ~\cite{pdg}.

\begin{figure}[hbtp]
\centering
\epsfig{width=0.45\textwidth,clip=true,file=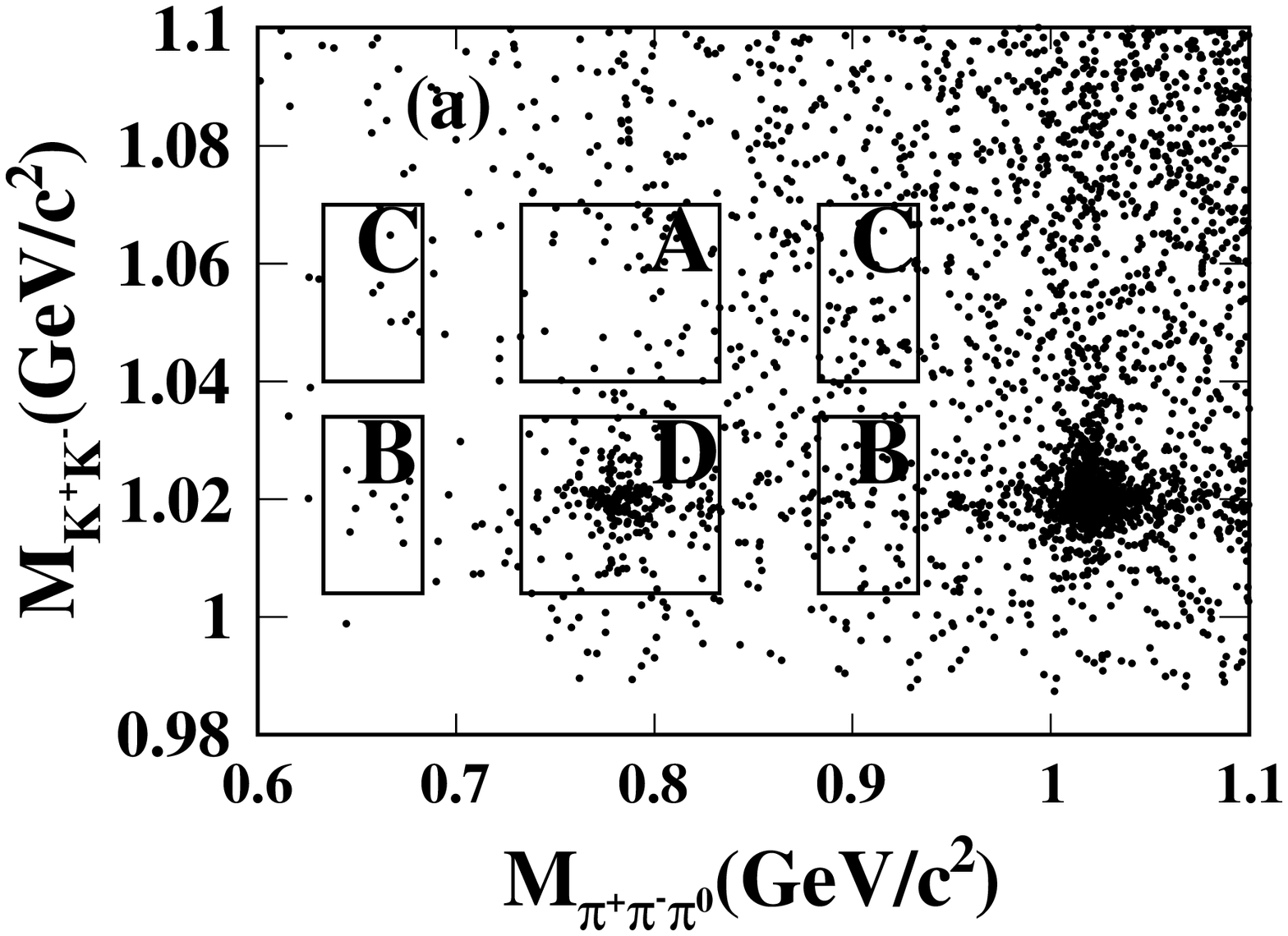}\\
\epsfig{width=0.45\textwidth,clip=true,file=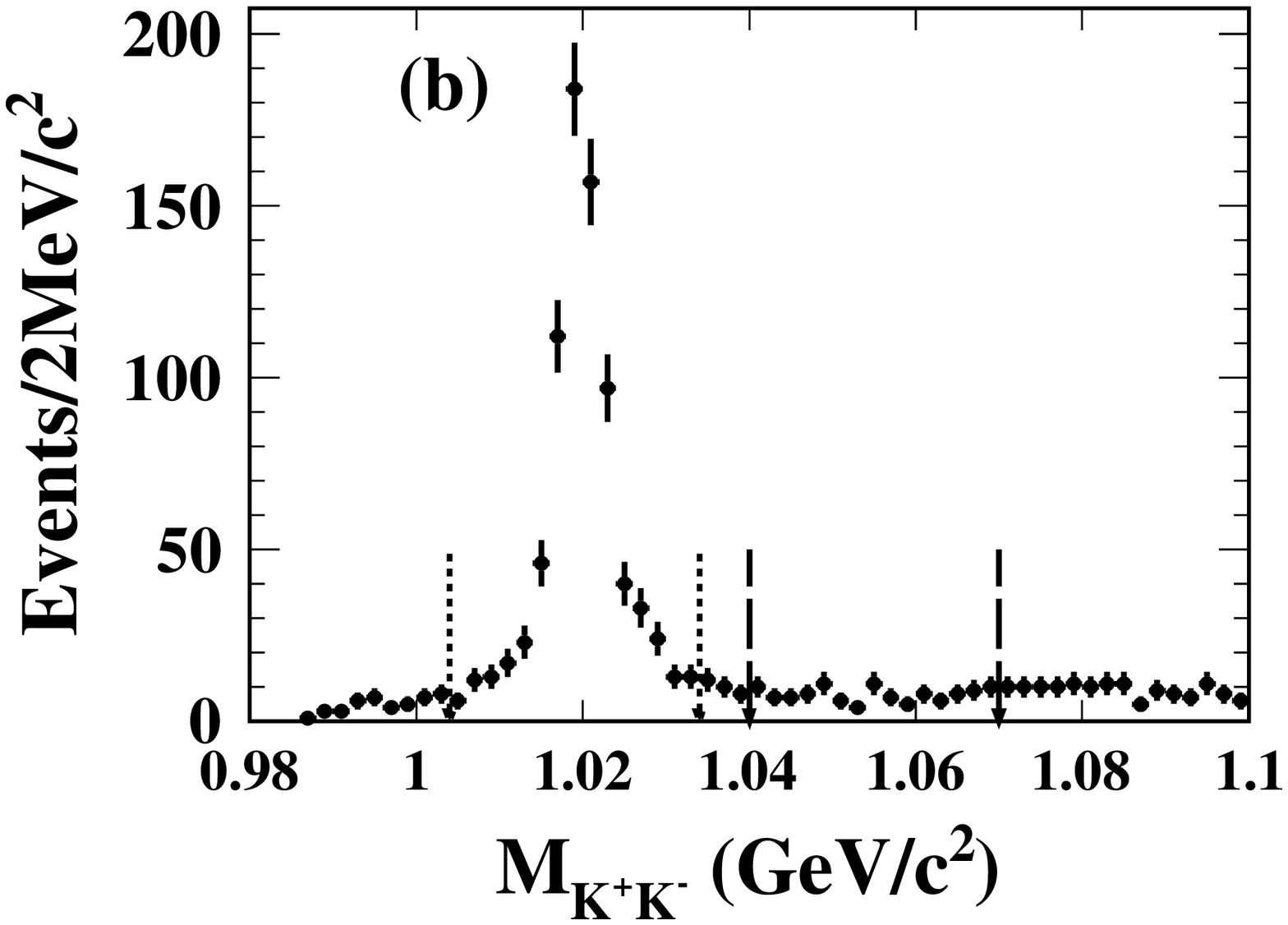}
\put(-120,120){$\phi$}\\
\epsfig{width=0.45\textwidth,clip=true,file=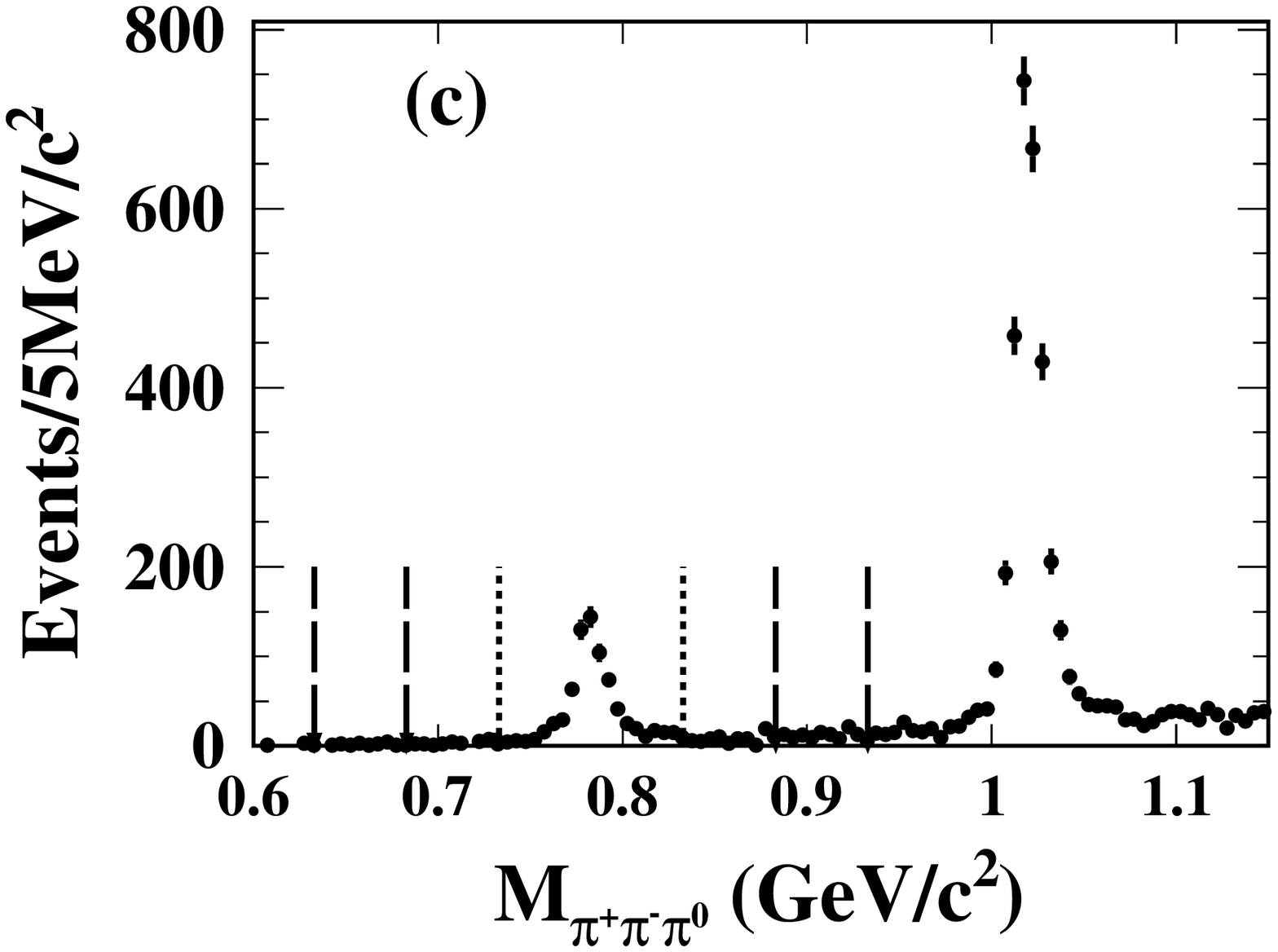}
\put(-130,60){$\omega$}
\put(-51,125){$\phi$}
\caption{(a) Scatter plot of $M_{K^+K^-}$ versus
  $M_{\pi^+\pi^-\pi^0}$ for events within the $\chi_{cJ}$ mass
  region.  The boxes indicate the sideband regions (labeled as A, B,
  and C) and signal region (labeled as D). (b) and (c) are the
  one-dimensional projection of the system recoiling against selected
  $\omega$ and $\phi$ candidates, respectively.
  The short-dashed arrows show the
  signal regions while long-dashed arrows show the sideband regions.}
\label{scatter}
\end{figure}

The $\pi^0$ candidates are selected
from the three $\GG$ combinations
as the pair with the
minimum $|M_{\GG}-M_{\pi^0}|$, where
$M_{\pi^0}$ is the nominal $\pi^0$ mass~\cite{pdg}.
Figure~\ref{scatter}(a) shows the plot of
the $K^+K^-$
versus  $\ppp$ invariant mass for the selected events
 in the $\chi_{cJ}$ signal region ([3.3, 3.6] GeV/$c^2$), and a clear accumulation at the $\omega$ and
$\phi$ masses is observed. The bottom-central
square $|M_{\pi^+\pi^-\pi^0}- M_{\omega}|<0.05$ GeV/c$^2$ and
$|M_{K^+K^-}- M_{\phi}|<0.015$ GeV/c$^2$ obtained by optimizing FOM, is taken as the $\omega\phi$
signal region (labeled as D), and the five squares around the signal
region are taken as the sideband regions (labeled as A, B and C), where
$M_{\ppp}~(M_{\kk})$ is the invariant mass of $\ppp~(\kk)$.
The $M_{\kk}$ distribution with $M_{\ppp}$ in
the $\omega$ signal region in Fig.~\ref{scatter}(b) shows
a clear $\phi$ peak.
Correspondingly, the $M_{\ppp}$ distribution
with $M_{\kk}$ within the $\phi$ signal region,
as shown in Fig.~\ref{scatter}(c), indicates clear $\omega$ and $\phi$ peaks.
The latter is from the decay $\chi_{cJ} \to \phi \phi \to \kk\ppp $.
Figure~\ref{mwf} shows
the invariant mass spectrum $M_{\kk\ppp}$ for events in the $\omega\phi$
sideband regions (sub-figures labeled A, B, and C) and the signal region
(sub-figure labeled D) with clear $\chi_{cJ}$ peaks in all plots.

Analysis of the $\psi(3686)$ inclusive MC sample indicates that the
peaking background in the $\chi_{cJ}$ signal region can be described
by the sideband events.  The data collected at $\sqrt{s}=$ 3.65 GeV
with an integrated luminosity of approximately $1/15$ of the $\psp$ data
are used to investigate non-resonant continuum
background. After the same event selection criteria are applied,
only a few
events survive, and they do not have any obvious enhancements in
the $\chi_{cJ}$ mass region.

\section{Signal Extraction}
\label{sec:results}
The number of the $\chi_{cJ}\to\omega\phi$ events is
determined by fitting the $M_{\kk\ppp}$ distributions within the $\omega\phi$ signal region (labeled as D in Fig.~\ref{scatter}(a)). The signal is described by the MC simulated shape convolved with a Gaussian function, which is used to account for the difference in
the $\chi_{cJ}$ mass and resolution between data and MC
simulation. The parameters of the Gaussian function are obtained using
the sample $\psi(3686)\to\gamma\chi_{cJ}\to\gamma\phi\phi\to\gamma\pi^+\pi^-\pi^0 K^+K^-$.

The peaking backgrounds from the $\chi_{cJ}\to\omega K^+K^-$, $\phi\pi^+\pi^-\pi^0$, and the non-resonant $K^+K^-\pi^+\pi^-\pi^0$  background are estimated using the
sideband regions labeled A, B, and C in Fig.~\ref{scatter}(a). The total peaking background contribution, $N_{\rm bkg}$, is the sum calculated as:
 \begin{eqnarray}
N_{\rm bkg}=N_{\rm bkg}^{\omega\kk}+N_{\rm bkg}^{\phi\ppp}+N_{\rm bkg}^{\rm n-r},
\label{eq2}
\end{eqnarray}
where $N_{\rm bkg}^{\omega\kk},~N_{\rm bkg}^{\phi\ppp}$, and $N_{\rm bkg}^{\rm n-r}$ are
numbers of the aforementioned peaking background contributions.
The contributions are determined using the following equations:
 \begin{eqnarray}
N_{\rm bkg}^{\omega\kk}&=&(N_{\rm A}-N_{\rm C}\cdot f_{\rm C\to A})\cdot f_{\rm A\to\rm D},\label{eq3}\\
N_{\rm bkg}^{\phi\ppp}&=&(N_{\rm B}-N_{\rm C}\cdot f_{\rm C\to B})\cdot f_{\rm B\to\rm D},\label{eq4}\\
N_{\rm bkg}^{\rm n-r}&=&N_{\rm C}\cdot f_{\rm C\to \rm D},\label{eq5}
\end{eqnarray}
where
$N_{\rm A}$, $N_{\rm B}$, and $N_{\rm C}$ are the numbers of the fitted $\chi_{cJ}$ events in the A, B, and C regions, respectively; $f_{\rm C\to A}$,
$f_{\rm C\to B}$, $f_{\rm C\to \rm D}$, $f_{\rm A\to \rm D}$, and $f_{\rm B\to \rm D}$ are the relative scaling factors for the
different regions. The factors are estimated using the corresponding MC simulation of $\chi_{cJ}\to K^+K^-\pi^+\pi^-\pi^0$, $\omega K^+K^-$, and $\phi\pi^+\pi^-\pi^0$. For example, $f_{\rm A\to D}$ is the ratio of the $\chicJ\to\omega K^+K^-$ yields between the $\rm D$ and $\rm A$ regions.

\begin{figure}[hbtp]
\centering
\epsfig{width=0.412\textwidth,clip=true,file=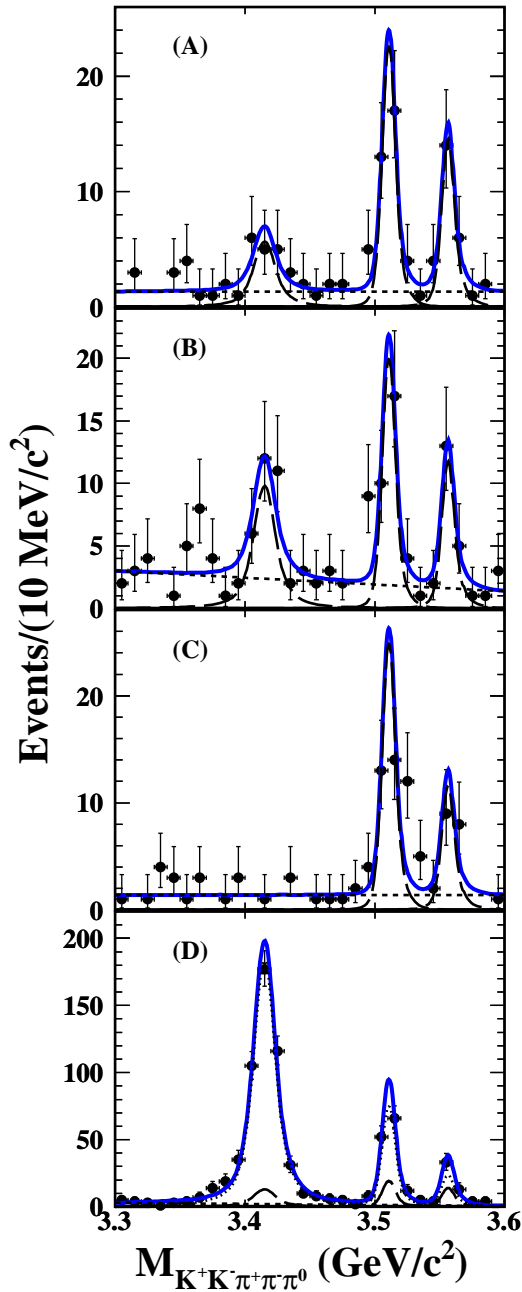}
\caption{Simultaneous fit to the $M_{\kk\ppp}$ distributions in
  the sidebands (A, B, and C) and the signal (D) regions.  The dots with error bars are data, the
  solid lines are the fit results, and the dotted lines
  represent the signal components.  The long-dashed line is background
  normalized using the simultaneous fit to the $\omega\phi$ sidebands, and
  the short-dashed line is the remaining background.}
\label{mwf}
\end{figure}

We perform a simultaneous unbinned maximum likelihood fit to the
$M_{\kk\ppp}$ distributions in the signal and sideband regions.
The result of the fit is
shown in Fig.~\ref{mwf}. The parameters of the Gaussian functions accounting for the difference between data and MC simulation are assumed to be the
same for the signal and sidebands. The
shape of the distributions outside the $\chicJ$ peaks is described by
 a polynomial function.
The statistical significance of the $\chi_{c1}~(\chi_{c2})$ signal is
determined by comparing the  $-2\rm ln{\cal{L}}$ value with the one from
the  fit  without  the $\chi_{c1}~(\chi_{c2})$ signal component,
and considering the change in the number of degrees of freedom.
The results are 12.3$\sigma$ and 4.8$\sigma$
for $\chi_{c1}$ and $\chi_{c2}$, respectively.
The extracted numbers of the  $\chicJ\to\omega\phi$ events
are given in Table~\ref{tab::chicjnb}.

\begin{table*}[!hbtp]
\begin{center}
  \caption{Number of signal events ($N_{\rm obs}^{\chi_{cJ}}$), detection efficiency ($\epsilon$), the product branching fraction ${\cal{B}}_{1}\times{\cal{B}}_{2}={\cal{B}}(\psi(3686)\to\gamma\chi_{cJ})\times{\cal{B}}(\chi_{cJ}\to\omega\phi)$, and the absolute branching fraction ${\cal{B}}(\chi_{cJ}\to\omega\phi)$. Here, the first uncertainty is statistical and the second systematic.}
\ \\
\begin{tabular}{c|c|c|c|c}
\hline
\hline
Mode      & $N_{\rm obs}^{\chi_{cJ}}$ & $\epsilon~(\%)$ &${\cal{B}}_{1}\times{\cal{B}}_{2}$ & ${\cal B}(\chi_{cJ}\to\omega\phi)$\\
\hline
$\chi_{c0}\to\omega\phi$   &  $486.3\pm24.5$  & $17.99\pm0.06$&$(13.83\pm0.70\pm1.01)\times10^{-6}$& $(13.84\pm0.70\pm1.08)\times10^{-5}$\\
$\chi_{c1}\to\omega\phi$   &  $104.7\pm12.1$  & $20.04\pm0.06$&$(2.67\pm0.31\pm0.27)\times10^{-6}$& $(2.80\pm0.32\pm0.30)\times10^{-5}$\\
$\chi_{c2}\to\omega\phi$   &  $32.9\pm8.3$  & $18.47\pm0.06$ & $(0.91\pm0.23\pm0.12)\times10^{-6}$& $(1.00\pm0.25\pm0.14)\times10^{-5}$\\
\hline
\hline
\end{tabular}
\label{tab::chicjnb}
\end{center}
\end{table*}


The product branching fractions, ${\cal{B}}(\psi(3686)\to\gamma\chi_{cJ})$ $\times{\cal{B}}(\chi_{cJ}\to\omega\phi)$ $={\cal{B}}_{1}\times{\cal{B}}_{2}$, are calculated as
{\small
\begin{eqnarray}
{\cal{B}}_{1}\times{\cal{B}}_{2}=\frac{N_{\rm obs}^{\chi_{cJ}}}{N_{\psp}\cdot{\cal B}(\omega)\cdot{\cal B}(\phi)\cdot{\cal B}(\pi^0)\cdot\epsilon},
\end{eqnarray}}

%

\noindent where $N_{\psp}$ is the number of $\psp$ events, ${\cal B}(\omega)$, ${\cal B}(\phi)$, and ${\cal B}(\pi^0)$
are the branching fractions of
$\omega\to\ppp$, $\phi\to\kk$, and $\pi^0\to\gamma\gamma$,
respectively~\cite{pdg}. The corresponding detection efficiencies,  $\epsilon$, are obtained from the MC simulations. The results for the product branching fractions are listed in
Table~\ref{tab::chicjnb}.

By using the world average values of ${\cal B}(\psp\to\gamma\chi_{cJ})$, the absolute branching fractions of $\chi_{cJ}\to \omega \phi$ are determined and also listed in Table~\ref{tab::chicjnb}.

\section{Systematic Uncertainties}
\label{sec:syserr}
\vspace{-0.4cm} The contribution of systematic effects on the product branching fractions 
from various sources is described in the following:
\begin{enumerate}
\item The tracking efficiency for
  $\pi$ and $K$ is investigated using control samples of
  $J/\psi\to\rho\pi$~\cite{prl107.092001} and $\psi(3686)\to\pi^{+}\pi^{-}K^{+}K^{-}$,
  respectively.  The difference in the
  efficiency for the track reconstruction between data and MC simulation is 1.0\% per
  pion and per kaon. Assuming they are all correlated, the uncertainty due to tracking efficiency
is 4\%.

\item The particle identification (PID) efficiency for kaons is investigated with control samples of $J/\psi\to K^{*}(892)^{0}K^{0}_{S} +c.c.$, and the systematic uncertainty
is determined to be 1\% per kaon track~\cite{prl107.092001}. In this analysis, only one of the two charged tracks is required to be identified as kaon. The bias on one of the two tracks being a kaon track will be much smaller than 1\%. Therefore, the uncertainty due to PID efficiency is negligible.

\item The uncertainty of the photon reconstruction efficiency is
  studied using $J/\psi\to\rho\pi$~\cite{prd81.052005}. The
  difference between data and MC simulation is found to be 1.0\% per
  photon, and the value 3\% is taken as the systematic uncertainty.

\item The uncertainty of MC generator comes from modeling $\psi(3686)\to\gamma\chi_{c1,2}$ and $\chicJ\to\omega\phi$ in MC simulation.

The uncertainty of assuming $\psi(3686)\to\gamma\chi_{c1,2}$ as pure E1 transition is studied by taking the higher-order multipile amplitudes contribution~\cite{M2E3} into account in the MC simulation. The resulting efficiency difference of 0.9\% for $\chi_{c1}$, and 0.5\% for $\chi_{c2}$, are taken as this systematic
  uncertainty.
The uncertainty of modeling $\chicJ\to\omega\phi$ is studied by changing the model from \textsc{helamp} to a pure phase space distribution.
  The resulting efficiency difference of 4.1\% for $\chi_{c0}$, 5.6\% for $\chi_{c1}$, and 1.3\% for $\chi_{c2}$, are taken as this systematic
  uncertainty.

  The total systematic uncertainties of MC generator are
obtained to be 4.1\% for $\chi_{c0}$, 5.7\% for $\chi_{c1}$, and 1.4\% for $\chi_{c2}$ by summing all individual contributions in quadrature, assuming two sources to be independent.

\item The uncertainty related to the $\pi^0$ mass window is
  studied by fitting the $\pi^0$ mass distribution of data and signal MC for the control
  sample $\psi(3686)\to\pi^{+}\pi^{-}\pi^{0}$. We obtain the $\pi^0$ detection efficiency,
  which is the ratio of the number of $\pi^0$ events selected with and without the $\pi^0$ mass
  window requirement, determined by integrating the fitted signal shape. The difference in the efficiency
  between data and MC simulation is 0.8\%.

\item The uncertainty related to the $M^{\rm recoil}_{\pi^+\pi^-}$ mass window requirement is
  studied with the control sample $\psi(3686)\to\pi^{+}\pi^{-}J/\psi, J/\psi\to\mu^+\mu^-$.
  We obtain the $M^{\rm recoil}_{\pi^+\pi^-}$ detection efficiency,
  which is the ratio of the number of events with and without the $M^{\rm recoil}_{\pi^+\pi^-}$ mass
  window. The difference in the efficiency between data and MC simulation is  0.4\%.

\item The systematic uncertainties associated with the 4C
  kinematic fit are studied with the track helix parameter correction method,
  as described in Ref.~\cite{prd87.012002}. In the standard analysis,
these corrections are applied. The difference
of the MC signal efficiencies with the uncorrected track
parameters are 1.5\%,  1.9\%,
  and 2.3\% for $\chi_{c0}$,  $\chi_{c1}$, and  $\chi_{c2}$ decays,
respectively. These values are taken
  as the uncertainties associated with the 4C kinematic fit.

\item The uncertainty related to the fitting comes from the fit range,
  $\omega$ and $\phi$ mass windows, sideband regions, and fitting function
  (including resolution and remaining backgrounds shape).
  \begin{enumerate}
  \item The uncertainty due to the fit range is estimated by changing the range by $\pm5$ MeV/c$^2$ in the mass spectrum, since the the non-resonant $K^+K^-\pi^+\pi^-\pi^0$  background shape is quite smooth according to the topology analysis with the inclusive MC sample. The largest differences for the
branching ratios are
  1.0\% for $\chi_{c0}$, 3.0\% for $\chi_{c1}$, and 10.0\% for
  $\chi_{c2}$. These numbers are assigned as the corresponding systematic
  uncertainties.


  \item The uncertainties associated with the $\phi$ and $\omega$ mass
  windows are estimated using two control samples,
  $\psp\to\gamma\chi_{cJ}$, $\chi_{cJ}\to\phi\phi\to(K^+K^-)~(\pi^+\pi^-\pi^0)$, and
  $\psp\to\gamma\chi_{cJ}$, $\chi_{cJ}\to\omega\omega\to2(\pi^+\pi^-\pi^0)$, respectively.  The
  efficiency for the $\omega$($\phi$) selection is obtained from the comparison of the
  $\omega$~($\phi$) yields determined from the
  $\pi^+\pi^-\pi^0$~($K^+K^-$) mass spectrum with and without the
  $\omega$~($\phi$) selection requirement. The difference in
  $\omega~(\phi)$-selection efficiency between data and MC simulation, 1.9\%~(0.6\%), is
  taken as the uncertainty of the $\omega~(\phi)$-mass window.

  \item The uncertainty due to background
estimates using the sidebands can be divided in two groups. One is due to
the sideband ranges, the other is due to contributions of various
intermediate states in $\chicJ\to\kk\ppp$ in the MC simulation used
to extract the scaling factors. The former can be estimated by changing the sideband range.
  By changing the mass region of $M_{\ppp}$ from $[0.633,0.683]/[0.883,0.933]$ to $[0.631,0.681]/[0.881,0.931]$ GeV/c$^2$, and the mass region of $M_{\kk}$ from
   $[1.04,1.07]$ to $[1.042,1.072]$ GeV/c$^{2}$, the differences of $\chi_{c0,1,2}$ signal yields are 0.6\%, 4.6\%, and 5.1\%, respectively.
  For the non-resonant $\chicJ\to\kk\ppp$, a phase space process was used.
The experimental distributions indicate the contribution of the
 intermediate states involving
 $K^{*}(892)$: $\chi_{cJ}\to K^{*0}\bar{K}^{*0}\pi^0$ and $\chi_{cJ}\to
  K^{*+}K^{-}\pi^+\pi^- +c.c.$. The corresponding MC distributions are mixed with the phase space model according to the ratios estimated from the fits
to data to recalculate the scale factors related to the region $\rm C$. The differences of
the $\chi_{c0,1,2}$ signal yields are 0.2\%, 2.0\%, and 4.2\%, respectively.
  The resulting differences
due to the two preceding effects
are found to be 0.6\%, 5.0\%, and 6.6\% for $\chi_{c0,1,2}$, respectively.

\item The systematic effects from the detector resolution difference between data and MC
  simulation are studied with the control sample $\psp\to\gamma\chi_{cJ}\to\phi\phi\to K^+ K^-\pi^+\pi^-\pi^0$. We
  change the difference by one standard deviation.
  No changes are found for the $\chi_{c0,1,2}$ signal yields
and these systematic uncertainties are neglected.

  \item The uncertainty from the non-$\chicJ$ background is estimated by
  changing the polynomial from first to second
  order in fitting $M_{\kk\ppp}$ mass spectrum. The differences in the final results are 0.4\%, 0.5\%, and 0.2\%, respectively.
  \end{enumerate}

\item The systematic uncertainties due to the branching
  fractions of $\omega\to\ppp$ and $\phi \to \kk$ are 0.8\% and 1.0\%, respectively~\cite{pdg}.
  Therefore, the uncertainties of the final results are 1.3\%. 

\item The number of $\psi(3686)$ events is estimated
  from the number of inclusive hadronic events, as described in
  Ref.~\cite{cpc42.023001}. The uncertainty of the total number of
  $\psi(3686)$ events is 0.6\%.
\end{enumerate}

Table~\ref{table:syserrA} summarizes the systematic uncertainties and
their sources for the product branching fractions. The total systematic uncertainties are
obtained by summing all individual contributions in quadrature, assuming all sources to be independent. For the uncertainties of absolute branching fractions $\chicJ\to\omega\phi$, the uncertainty arising from $\psi(3686)\to$ $\gamma\chi_{cJ}$ transition rate is added.

\begin{table}[hbtp]
\begin{center}
  \caption{Relative contributions to systematic uncertainties in measuring the product branching fraction of
    ${\cal{B}}_{1}\times {\cal{B}}_{2}={\cal{B}}(\psi(3686)\to\gamma\chi_{cJ})\times{\cal{B}}(\chi_{cJ}\to\omega\phi)$ (in units of \%).}
\ \\
\begin{tabular}{c|c|c|c}
\hline
\hline
Source  & $\chi_{c0}$ & $\chi_{c1}$  & $\chi_{c2}$ \\
\hline
Tracking efficiency        &  4.0 & 4.0 & 4.0  \\
PID efficiency             &  negligible & negligible & negligible \\
Photon efficiency          &  3.0 & 3.0 & 3.0  \\
MC generator  &  4.1 & 5.7 & 1.4  \\
$\pi^0$ mass window         &  0.8 & 0.8 & 0.8  \\
$M^{\rm recoil}_{\pi^+\pi^-}$ mass window         &  0.4 & 0.4 & 0.4  \\
Kinematic fit              &  1.5 & 1.9 & 2.3  \\
Fit range &  1.0 & 3.0 & 10.0  \\
$\omega$ mass window &  1.9 &1.9 & 1.9  \\
$\phi$ mass window &  0.6 &0.6 & 0.6  \\
Sidebands &  0.6 & 5.0 & 6.6  \\
Resolution &  negligible & negligible & negligible  \\
Remaining background shape &  0.4 & 0.5 & 0.2  \\
Intermediate state  &  1.3 & 1.3  & 1.3   \\
$N_{\psp}$  &  0.6 & 0.6 & 0.6  \\
\hline
Total                      &  7.2&10.2&13.7 \\
\hline
\hline
\end{tabular}
\label{table:syserrA}
\end{center}
\end{table}

\section{results and discussion}
\label{sec:summary}
\vspace{-0.4cm}

Using the data sample of
$(448.1\pm2.9)\times10^6$ $\psi(3686)$ events collected with the
BESIII detector,
we present the improved measurement of the doubly OZI suppressed
decays $\chi_{cJ}\to\omega\phi$. The decay $\chi_{c1}\to\omega\phi$ is observed for the first
time with a 12.3$\sigma$ statistical significance and the branching fraction of $\chi_{c0}\to\omega\phi$ is measured with improved precision.
We also observe strong evidence for $\chi_{c2}\to\omega\phi$ at a statistical significance of 4.8$\sigma$.
The product branching fractions, ${\cal{B}}(\psi(3686)\to\gamma\chi_{c0,1,2})\times{\cal{B}}(\chi_{c0,1,2}\to\omega\phi)$,
and the absolute branching fractions, ${\cal{B}}(\chi_{c0,1,2}\to\omega\phi)$, are determined as listed in Table~\ref{tab::chicjnb}.
In addition, using the branching fractions of $\chi_{c1}\to
\omega\omega, \phi\phi$ from Ref.~\cite{prl107.092001}, the ratios
${\cal{B}}(\chi_{c1}\to\omega\phi)/{\cal{B}}(\chi_{
  c1}\to\omega\omega)$ and ${\cal{B}}(\chi_{
  c1}\to\omega\phi)/{\cal{B}}(\chi_{c1}\to\phi\phi)$
of $(4.67\pm 0.78)\times10^{-2}$ and $(5.60\pm
1.01)\times10^{-2}$ are obtained, respectively. Here, the common
systematic uncertainties in the two measurements cancel in
the ratio.
These ratios are one order of magnitude larger than the
theoretical predictions~\cite{prd81.074006}.
These measurements will be helpful in clarifying the influence of the long-distance contributions in this energy region,
understanding the theoretical dilemma surrounding the OZI and HS rules, and
checking mesonic loop contributions and the $\omega-\phi$ mixing effect.

\begin{acknowledgements}
\label{sec:acknowledgement}
\vspace{-0.4cm}

The BESIII collaboration thanks the staff of BEPCII and the IHEP computing center for their strong support. This work is supported in part by the National Key Basic Research Program of China under Contract No. 2015CB856700; National Natural Science Foundation of China (NSFC) under Contracts Nos. 11335008, 11425524, 11625523, 11635010, 11735014, 11605042; the Chinese Academy of Sciences (CAS) Large-Scale Scientific Facility Program; the CAS Center for Excellence in Particle Physics (CCEPP); Joint Large-Scale Scientific Facility Funds of the NSFC and CAS under Contracts Nos. U1632109, U1532257, U1532258, U1732263; CAS Key Research Program of Frontier Sciences under Contracts Nos. QYZDJ-SSW-SLH003, QYZDJ-SSW-SLH040; 100 Talents Program of CAS; CAS Open Research Program of Large Research Infrastructures under Contract No. 1G2017IHEPKFYJ01; INPAC and Shanghai Key Laboratory for Particle Physics and Cosmology; German Research Foundation DFG under Contracts Nos. Collaborative Research Center CRC 1044, FOR 2359; Istituto Nazionale di Fisica Nucleare, Italy; Koninklijke Nederlandse Akademie van Wetenschappen (KNAW) under Contract No. 530-4CDP03; Ministry of Development of Turkey under Contract No. DPT2006K-120470; National Science and Technology fund; The Swedish Research Council; U. S. Department of Energy under Contracts Nos. DE-FG02-05ER41374, DE-SC-0010118, DE-SC-0010504, DE-SC-0012069; University of Groningen (RuG) and the Helmholtzzentrum fuer Schwerionenforschung GmbH (GSI), Darmstadt; China Postdoctoral Science Foundation under Contracts Nos. 2017M622347, Post-doctoral research start-up fees of Henan Province under Contract No. 2017SBH005 and Ph.D research start-up fees of Henan Normal University under Contract No. qd16164; Program for Innovative Research Team in University of Henan Province (Grant No.19IRTSTHN018).

\end{acknowledgements}

\end{document}